\documentclass[12pt,preprint]{aastex}

\begin{document}


\slugcomment{accepted by ApJL: March 8, 2012}

\title{
A Resolved Millimeter Emission Belt in the AU~Mic Debris Disk
}

\author{
David J. Wilner\altaffilmark{1},
Sean M. Andrews\altaffilmark{1},
Meredith A. MacGregor\altaffilmark{1},
A. Meredith Hughes\altaffilmark{2}
}
\altaffiltext{1}{Harvard-Smithsonian Center for Astrophysics, 60 Garden Street, Cambridge, MA 02138}
\altaffiltext{2}{Department of Astronomy, 601 Campbell Hall, University of 
California, Berkeley, CA 94720} 

\begin{abstract}
We present imaging observations at 1.3~millimeters of the debris disk 
surrounding the nearby M-type flare star AU~Mic with beam size $3''$ (30~AU)
from the Submillimeter Array. 
These data reveal a belt of thermal dust emission surrounding the star with the 
same edge-on geometry as the more extended scattered light disk detected at 
optical wavelengths. Simple modeling indicates a central radius of $\sim35$~AU 
for the emission belt. This location is consistent with the reservoir of 
planetesimals previously invoked to explain the shape of the scattered light 
surface brightness profile through size-dependent dust dynamics. The 
identification of this belt further strengthens the kinship between the debris 
disks around AU~Mic and its more massive sister star $\beta$~Pic, members of 
the same $\sim10$~Myr-old moving group.
\end{abstract}

\keywords{
circumstellar matter ---
planet-disk interactions---
stars: individual (AU~Microscopii) ---
submillimeter: planetary systems
}

\section{Introduction}
AU~Mic is a nearby \citep[$9.91\pm 0.10$~pc,][]{van07} M1Ve flare star 
\citep{tor06} with the young age of $12^{+8}_{-4}$~Myr \citep{zuc01}, a key 
epoch in the formation of planetary systems. The star is surrounded by a nearly 
edge-on circumstellar disk discovered in coronographic images of scattered 
starlight that extends to a radius of at least 210~AU \citep{kal04}. Since the 
small grains responsible for scattering should be removed rapidly by stellar 
pressure forces, the disk is thought to consist of the collisional debris of 
unseen planetesimals experiencing ongoing impacts \citep[for recent reviews of 
the debris disk phenomenon, see][]{wya08,kri10}. Because of its proximity, the 
AU~Mic debris disk has become one of the best studied examples at optical and 
infrared wavelengths, including detailed imagery of both intensity and 
polarization from the {\em Hubble Space Telescope} \citep{kri05,gra07}. 

Many observational properties of the AU~Mic debris disk bear striking 
similarities to the archetype debris disk surrounding $\beta$~Pic, also viewed 
nearly edge-on and located in the same young moving group. In particular, the 
midplane optical surface brightness profiles of these disks are remarkably 
alike, with shallow inner slopes ($\sim 1/r^{1-2}$) that steepen substantially 
($\sim 1/r^{4-5}$) in the outer regions, near 35~AU and 100~AU for AU~Mic and 
$\beta$~Pic, respectively \citep{kri05,fit07,hea00,gol06}. These similarities 
inspired the development of a unified framework for debris disks based on a 
localized belt of planetesimals, or ``birth ring'', that produces dust in a 
collisional cascade \citep{str06,aug06}. The smallest grains are blown out from 
the belt by radiation and winds from the star, larger grains are launched into 
eccentric orbits with the same periastron as the belt, and the largest grains, 
which are minimally affected by non-gravitational forces, do not travel far 
from the belt before being ground down. The grains are therefore segregated 
according to their size, which gives rise to the characteristic scattered light 
profile. For these dynamics to prevail, the disks must be substantially free of 
gas \citep{the05}, a property confirmed by measurements at ultraviolet 
wavelengths \citep{rob05,fra07}. 

A ``birth ring'', if present, would remain hidden in optical and infrared 
images dominated by small grains that populate an extended halo 
\citep[e.g.,][]{su05}. By contrast, observations at (sub)millimeter wavelengths 
highlight thermal emission from the largest grains and hence the location of 
the dust-producing planetesimals \citep{wya06}. To date, the best case for a 
``birth ring'' comes from millimeter observations of $\beta$~Pic \citep{wil11}. 
While the optical disk of $\beta$~Pic extends more than $1000$~AU from the 
star, the millimeter imaging reveals a much more compact belt of emission at 
$\sim95$~AU radius. This millimeter emission belt coincides with the change in 
the slope of the optical surface brightness, which in the models marks the 
outer extent of the colliding planetesimals. 

Previous (sub)millimeter-wave observations of AU Mic did not have sufficient 
angular resolution to reveal much about the emission morphology. A detection at 
850~$\mu$m using JCMT/SCUBA ($14''$ beam) indicated a reservoir of cold dust 
with mass $\sim0.01$~M$_{\oplus}$, but did not resolve any structure 
\citep{liu04b}. Subsequent observations at 350~$\mu$m using CSO/SHARC~II 
($10''$ beam) marginally resolved an orientation compatible with the scattered 
light, but were otherwise limited \citep{che05}. In this {\em Letter}, we 
present imaging observations of AU Mic at 1.3~millimeters using the 
Submillimeter Array (SMA)\footnote{The Submillimeter Array is a joint project 
between the Smithsonian Astrophysical Observatory and the Academica Sinica 
Institute of Astronomy and Astrophysics and is funded by the Smithsonian 
Institution and the Academica Sinica.} that obtain $3''$ resolution and provide 
evidence for a planetesimal belt.

\section{Observations} \label{sec:obs}
We observed AU~Mic with the SMA \citep{ho04} on Mauna Kea, Hawaii at 
1.3~millimeters wavelength using the compact and extended configurations of the 
array. Table~\ref{tab:obs} provides basic information about these observations, 
including the observing dates, baseline lengths, and atmospheric opacities. The 
weather conditions were good for all of these observations and best for the two 
compact configuration tracks when only 6 of the 8 array antennas were 
available. The phase center was $\alpha = 20^h45^m09\fs53$, $\delta = 
-31\degr20\arcmin27\farcs2$ (J2000), offset by $\sim5\farcs3$ from the location 
of the star. The $\sim54''$ (FWHM) field of view is set by the primary beam 
size of the 6~meter diameter array antennas. The total bandwidth available was 
8~GHz derived from two sidebands spanning $\pm4$~to~$8$~GHz from the LO 
frequency. Time dependent complex gains were calibrated using observations of 
two quasars, J2101-295 (3.9 degrees away) and J1924-292 (17.4 degrees away),
interleaved with observations of AU Mic in a 15~minute cycle. The passband 
shape for each track was calibrated using available bright sources, mainly 
J1924-292. The absolute flux scale was set with an accuracy of $\sim10\%$ using 
observations of Callisto or Ganymede in each track. All of the calibration 
steps were performed using the IDL based MIR software, and imaging and 
deconvolution were done with standard routines in the MIRIAD package. We made a 
series of images with a wide range of visibility weighting schemes to explore 
compromises between higher angular resolution and better surface brightness 
sensitivity. 

\begin{deluxetable}{lcccc}
\tablecaption{
Submillimeter Array Observations of AU Mic
\label{tab:obs}
}
\tablewidth{0pt}
\tablehead{
\colhead{Observation} & \colhead{2011 Jul 31} & \colhead{2011 Sep 26} & 
                      \colhead{2011 Sep 27} & \colhead{2011 Oct 25} 
}
\startdata
Array Configuration & Extended & Compact & Compact & Compact \\
Number of Antennas & 8 & 6 & 6 & 7 \\
Baseline Lengths (m) & 10--189 & 8--68 & 8--68 & 8--68 \\
LO frequency (GHz) & 235.6 & 225.4 & 225.4 & 235.6 \\
225 GHz atm. opacity\tablenotemark{a} & 0.10  & 0.10--0.06 & 0.06 & 0.08 \\
\enddata
\tablenotetext{a}{Measured at the nearby Caltech Submillimeter Observatory.}
\end{deluxetable}

\section{Results and Analysis} \label{sec:result}
\subsection{1.3 Millimeter Emission} \label{sec:continuum}
Figure~\ref{fig:image} shows a contour image of the 1.3~millimeter emission 
overlaid on a {\it Hubble Space Telescope}/ACS coronographic image of optical 
scattered light (F606W filter) from \citet{gra07}. The synthesized beam size 
for this 1.3~millimeter image is $3\farcs3 \times 2\farcs6$ ($33\times26$~AU), 
position angle $-12{\degr}$, obtained with natural weighting and an elliptical 
Gaussian taper $1\farcs5 \times 1\farcs0$ (FWHM) oriented east-west to make the 
beam shape more circular. The rms noise in this image is 0.40~mJy~beam$^{-1}$, 
and the individual peaks have a signal-to-noise $\gtrsim7$. The star symbol 
marks the stellar position corrected for proper motion, offset by $(3\farcs20, 
-4\farcs25)$ from the phase center. The image shows a resolved band of 1.3 
millimeter emission that extends approximately symmetrically from the stellar 
position to the southeast and northwest. By fitting a line to the positions of 
the two peaks in the image, we estimate the position angle of the emission 
structure to be $130\pm 2\degr$, in excellent agreement with the disk 
orientation inferred from scattered light observations \citep{liu04a}. From its 
position angle and double peaked morphology, we identify the emission structure 
as a limb-brightened dust belt. Note that thermal emission from the stellar 
photosphere contributes only $0.06$~mJy at this wavelength, and even the 
strongest synchrotron radio flares from stellar activity should remain well 
below the noise \citep{kun87}. The total flux density obtained by integrating 
over a box that surrounds the emission is $8.5\pm2$~mJy, where the uncertainty 
is estimated conservatively by evaluating boxes of the same size in 
emission-free regions of the image. No other significant features are detected 
in the field of view.

An extrapolation of the measurements at shorter wavelengths indicates that 
missing flux due to the spatial filtering by the SMA is not a significant issue 
for these observations. The spectral index between 350~$\mu$m 
\citep{che05} and 850~$\mu$m 
\citep{liu04b} is $1.7^{+0.3}_{-0.5}$, which predicts 
$7.1^{+2.0}_{-1.2}$~mJy at 1.3~millimeters, in good agreement with the SMA 
measurement. This shallow index is consistent, within the uncertainties, with 
values near 2.0 determined for a sample of debris disks by \citet{gas11}, 
notably for stars later than F-type where dust temperatures are so low that the 
Rayleigh-Jeans approximation in this part of the spectrum is invalid.  We 
conclude that the compact structure detected by the SMA accounts for all of the 
1.3~millimeter emission in the AU~Mic system.

\subsection{A Simple Disk Model} \label{sec:models}
To characterize the millimeter emission structure, we use a simple parametric 
disk model, following closely the method used by \citet{wil11} for analyzing 
similar observations of the $\beta$~Pic disk. We assume the emission arises 
from a geometrically thin, axisymmetric belt with a radial surface brightness 
profile given by $I\propto f(r)r^{-q}$. This functional form is intended to 
capture the essence of low optical depth emission from a disk with a spatially 
invariant dust emissivity, surface density profile $f(r)$, and temperature 
profile that falls off with radius as a power-law. We fix the power-law index 
$q=0.5$ to approximate radiative heating from the central star, and we consider 
two shapes for $f(r)$: (1) an annulus with power law slope, $f(r) \propto 
r^{-p}$ for $R \pm \Delta R/2$, $p \in {0,1}$, and (2) a Gaussian, $f(r) 
\propto \exp{[-((r-R)/ \sqrt{2} \Delta R)^2]}$. In each case, the model is 
described by three parameters: a center radius $R$, width $\Delta R$, and total 
flux density $F = \int I d\Omega$. We fix the disk inclination at $89\fdg5$ and 
position angle at $130\degr$, as determined from scattered light data 
\citep{kri05}, and place the origin within $0\farcs1$ of the nominal stellar 
position (within the SMA astrometric accuracy). Note that the vertical 
thickness of the millimeter grain population is expected to be $<0\farcs25$ 
\citep{the09}, negligible compared to the resolution of the data. The utility 
of this simple modeling approach is to provide basic constraints while avoiding 
the multitudinous assumptions about grain properties, dynamics, and radiative 
transfer required in more sophisticated treatments. 

\begin{figure}[ht!]
\begin{center}
\includegraphics[scale=1.00,angle=0]{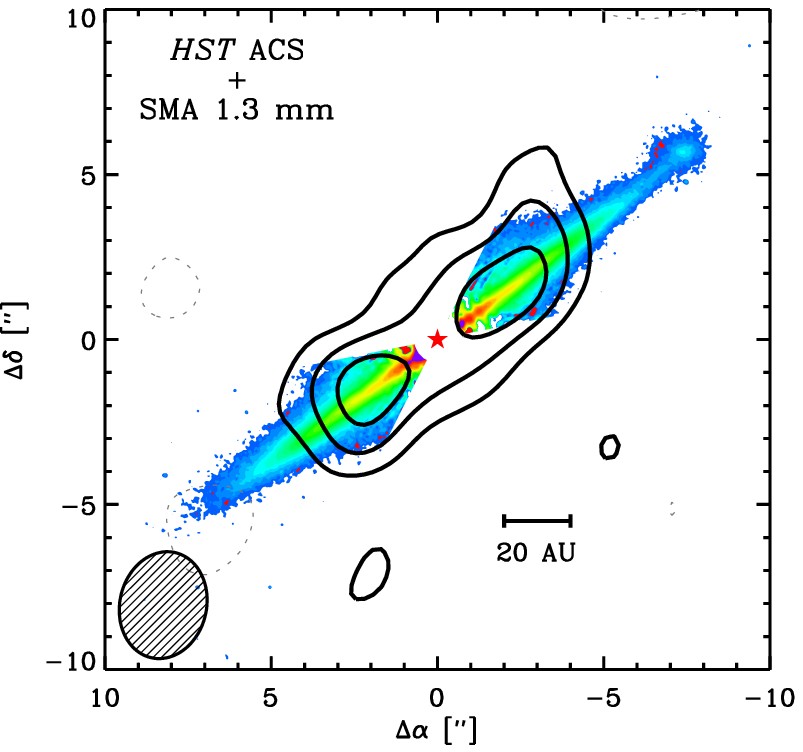}
\figcaption{
SMA image of the 1.3 millimeter continuum emission from AU~Mic, overlaid on an 
image of optical scattered light from the {\it Hubble Space Telescope} 
\citep{gra07}. The contour levels are $-2,2,4,6\times0.40$~mJy (the rms noise 
level); negative contours are dotted. The ellipse in the lower left corner 
represents the $3\farcs3 \times 2\farcs6$ (FWHM) synthesized beam size. The 
star symbol indicates the location of the stellar photosphere. 
\label{fig:image}}
\end{center}
\end{figure}

\begin{figure}[ht!]
\begin{center}
\includegraphics[scale=0.9,angle=0]{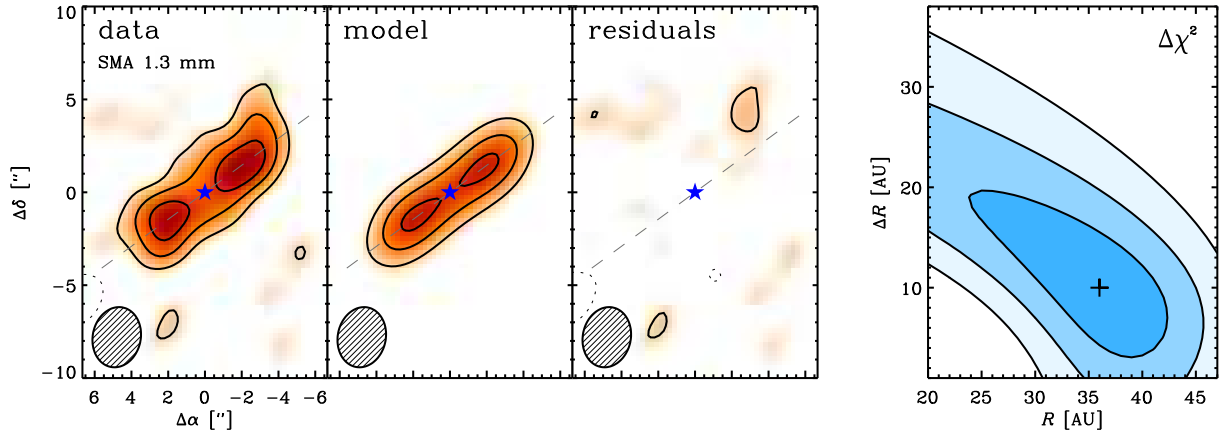}
\figcaption{
{\em Left panels:} SMA image of the 1.3 millimeter emission from AU~Mic 
together with images of the best-fit axisymmetric belt model and residuals (see 
text). The contour levels and beam size are the same as in 
Figure~\ref{fig:image}. The dashed line indicates the $130{\degr}$ position 
angle of the scattered light disk. {\em Right panel:} The $\chi^2$ surface for 
the model emission belt center and width parameters, with contours at 
$1,2,3\sigma$. The cross marks the best-fit model.
\label{fig:model}}
\end{center}
\end{figure}

To estimate the model parameters and their uncertainties, we calculate 
three-dimensional grids spanning appropriate ranges of the parameter values. 
For the Gaussian model, the grid covers $20 < R < 47$~AU and $1 < \Delta R < 
40$~AU in 1~AU steps, and $5 < F < 11$~mJy in 0.2~mJy steps.  For each position 
in a model grid, we compute a set of synthetic SMA visibilities and compare 
directly to the data with a $\chi^2$ value (the sum of real and imaginary 
components over all spatial frequencies) that uses natural weights modified by 
the Gaussian taper used for the image in Figure~\ref{fig:image}. The rightmost 
panel of Figure~\ref{fig:model} shows the $\chi^2$ surface in the $(R,\Delta 
R)$ plane for the Gaussian model after marginalizing over $F$. The best-fit 
model is marked by a cross, and the contours delineate (marginalized) $1 
\sigma$ intervals in $\Delta \chi^2$. The best-fit center radius is $R = 
36^{+7}_{-16}$~AU, with corresponding width $\Delta R = 10^{+13}_{-8}$~AU (or 
FWHM = $2\sqrt{2\ln2} \Delta R = 23^{+31}_{-19}$~AU) and flux density $F = 8.0 
\pm 1.2$~mJy. The three lefthand panels of Figure~\ref{fig:model} show the 
original 1.3~millimeter image from Figure~\ref{fig:image}, the Gaussian model 
image obtained using the best-fit parameter values and the same visibility 
weighting scheme, and the imaged residuals (where data$-$model subtraction is 
conducted on the visibilities).  This comparison shows clearly that the 
best-fit model reproduces the main features of the image and leaves no 
systematic residuals. The power-law models for $f(r)$ give similar best-fit 
parameter values: for $p=0$, we find $R=35\pm6$~AU and $\Delta R = 
38^{+26}_{-18}$~AU (where the upper limit is set by the computational grid 
boundary). For $p=1$, the best-fit locus is shifted to roughly $10\%$ larger 
values of $R$, a manifestation of the degeneracy between the radial gradient 
and emission extent. All of these best-fit models reproduce the observed SMA 
visibilities equally well (i.e., have effectively the same minimum $\chi^2$ 
values).  

The modest signal-to-noise of the observations preclude placing tighter 
constraints on the model parameters. Nonetheless, it is reassuring that 
consistent results are obtained from the various assumptions about the emission 
morphology. Each gives a similar center radius, as well as an effectively 
cleared central cavity with size compatible with that inferred from scattered 
light \citep[$\sim12$~AU][]{kri05} and the lack of mid-infrared emission 
\citep[$\sim17$~AU][]{liu04b}. The best-fit Gaussian model reproduces better 
the contrast of the maxima to the extended emission of the disk in the image 
plane, but there is no clear preference for any of the functional forms. Of
course, none of these {\em ad hoc} models are perfect, and more sensitive data 
will be needed to constrain further the details of the emission distribution.

\section{Discussion} \label{sec:discussion}
We have spatially resolved the thermal emission at 1.3 millimeters from the 
AU~Mic debris disk, thereby revealing the distribution of its millimeter grain 
population. These large grains are expected to have dynamics similar to the 
unseen dust-producing planetesimals, and the emission at this long wavelength 
provides a direct link to their spatial distribution. 

The overall structure of the AU~Mic millimeter emission is reproduced well by a 
belt surrounding the star centered near a radius of $35$~AU. The inferred 
location of this emission belt coincides with the region over which the 
scattered light brightness profile of the disk steepens markedly, the feature 
that provided impetus for developing the ``birth ring'' model. Like AU~Mic's 
more massive sister star $\beta$~Pic \citep{wil11}, the resolved 
multi-wavelength data are consistent with a scenario where destructive 
collisions of planetesimals within the belt create grains with a spectrum of 
sizes, and the effects of size-dependent dust dynamics generate the millimeter 
emission and spatially extended scattered light nebula. For $\beta$~Pic, the 
intense stellar radiation pressure can account for the large halo of small 
grains in the outer disk.  But for AU~Mic, which is about two orders of 
magnitude less luminous, the much weaker stellar radiation field is apparently 
augmented by a stellar wind to expel small grains \citep{str06,aug06}.  

Though the fractional radial extent of the AU~Mic millimeter emission is not 
strongly constrained, the best-fit models suggest that it could be broad, 
$\Delta R/R \sim 1$. For $\beta$~Pic, the millimeter emission is characterized 
by $\Delta R/R < 0.5$, with the upper end favored for the width of the 
underlying dust-producing planetesimal belt from detailed models of the knee of 
the scattered light profile \citep{hah10}. Such broad belts are not unusual 
features of debris disks, in particular among the few resolved at millimeter 
wavelengths \citep{hug11}. An inner cavity devoid of dust is plausibly 
maintained by the combination of grinding collisions in the belt and clearing 
by stellar radiation forces, though a tenable alternative is that the inner 
regions are swept clear by a planet. Planets seem to be the most viable 
explanation for the millimeter cavities common to the gas-rich transition disks 
around younger, pre-main-sequence stars \citep{and11}. While a planet has been 
imaged to orbit within the cavity in the disk around $\beta$~Pic \citep{lag10},
direct evidence for planets in the AU Mic system remains elusive, both from 
high contrast imaging observations \citep{mas05,met05} and from transit 
searches \citep{heb07}. The outer boundary of the emission belts may reflect 
the outer extent of successful planetesimal formation, or it may simply 
mark the edge of the region where the planetesimal stirring mechanism has 
been successful at initiating dust-producing collisions, as in the models of 
\citet{ken04}. 

These new millimeter observations of AU~Mic show the global structure of large 
grains in its debris disk, but they do not have sufficient resolution and 
sensitivity to reveal any significant departures from axisymmetry or 
substructure that might result from gravitational interactions with unseen 
planets. The higher resolution images of scattered light clearly show several 
radial and vertical inhomogeneities at subarcsecond scales \citep[features A-E, 
see][]{liu04a,met05,fit07}. If these substructures have counterparts at 
millimeter wavelengths, then future observations with the Atacama Large 
Millimeter/Submillimeter Array (ALMA) may be able to detect them, pinpoint the 
location of parent bodies, and constrain their origins. 

\acknowledgments{
We thank James Graham for providing the {\it Hubble Space Telescope} image in 
Figure 1.  A. M. H. is supported by a fellowship from the Miller Institute for 
Basic Research in Science.
}

{\it Facility:} \facility{Submillimeter Array}

\end{document}